\documentstyle[preprint,floats,epsf,eqsecnum,aps,tighten]{revtex} 
\begin{document}
\draft

\preprint{
 \parbox{1.5in}{\leftline{JLAB-THY-97-04}
                \leftline{WM-97-102}                  
                \leftline{nucl-th/9702014} }  }  
\title{ Normalization of the covariant three-body bound state vertex function }
\author{J.~Adam, Jr.,$^{1,*}$ Franz Gross,$^{2,1}$ \c{C}etin
\c{S}avkl\i,$^{2}$  and J. W. Van Orden$^{3,1}$ }
\address{$^1$Jefferson Lab,
12000 Jefferson Avenue, Newport News, VA 23606\\
$^2$Department of Physics, College of William and Mary,
Williamsburg, Virginia 23185\\
$^3$Department of Physics, Old Dominion University, Norfolk, VA 23529}

\date{\today}

\maketitle

\begin{abstract}

The normalization condition for the
relativistic three nucleon  Bethe-Salpeter and Gross bound state
vertex functions is derived, for the first time, directly from the
three body wave equations.  It is also shown that the
relativistic normalization condition for the two body Gross
bound state vertex function is identical to the requirement that
the bound state charge be  conserved, proving that charge is
automatically conserved by this equation.

\end{abstract}

\def\s{\ \! / \! \! \! \!}

%\centerline{\hbox{\epsfysize=3in \epsfbox{figure1.eps}}}

%%\narrowtext
\widetext
\section{Introduction}

The use of vertex functions derived from the covariant Bethe-Salpeter (BS)
equation\cite{BS,Tay} or the covariant spectator (or Gross)
equation\cite{Gross,Gross3b} to calculate matrix elements
involving bound states requires  that the normalization
of these vertex functions be determined. Although this
may  seem like a trivial problem, the complicated forms
of these equations, along  with the generally nonlocal
nature of their interaction kernels makes it more than
trivial.  In fact  we know of only one derivation of the
normalization condition for the three-body BS bound state
vertex functions\cite{3BR}, and this derivation does not
seem to show that the normalization condition follows
simply from application of the wave equation to the
description of scattering in the region of the bound
state, and does not obtain the normalization condition for
the Faddeev subvertex functions.  The three body spectator
equations have only recently been derived in their final
form\cite{SG} and this paper includes the first derivation of
the normalization condition of the three body spectator vertex
functions.  Normalization conditions for the two-body BS  
and Gross equations have been previously obtained by
many people (see, e.g., \cite{Lou,GVOH,deutletter}), 
but this paper includes a first demonstration
that charge is automatically conserved by the spectator
equations.

Following this introductory discussion the paper begins, in Sec.~II, with
a derivation of the normalization condition for both the two-body BS and
Gross vertex functions.  These results, which are not new, are presented
in order to introduce the techniques which will be used later in the
derivation of the three-body conditions. At the conclusion of this section
we prove, for the first time, that the normalization of the two-body Gross
vertex function is identical to the requirement that the charge of the
bound state, as defined by the charge operator introduced by Gross and
Riska\cite{GR}, is equal to the sum of the charges of its constituents.
This shows that charge is automatically conserved by the spectator theory.

The third section contains the new derivations of the
normalization conditions for the three-body
Bethe-Salpeter and Gross vertex functions
\cite{Gross3b}, which closely follow the approach
used in the two-body case.  A discussion of the Feynman
rules used in this paper and the conventions used in
symmetrizing the scattering matrices is presented in   the
Appendix.

\section{Normalization of Relativistic Two-Body Vertex Functions}

The derivations of the normalization condition for two and three-body
vertex functions are very similar. In this section we start with the simpler
two-body case, and then generalize the derivation to the three body case
in Sec.~III.

\subsection{Normalization of the BS two-body vertex function}

The two-body Bethe-Salpeter equation for the scattering matrix  ${\cal M}$
is represented by the Feynman diagrams in Fig.~\ref{2_body_scatt_mat}.
Using the Feynman rules described in the Appendix, this
corresponds to
\begin{eqnarray}
{\cal M}=&&V-VG_{BS}{\cal M}\label{2bodyeq01} \\
=&&V-{\cal M}G_{BS}V\, , \label{2bodyeq02}
\end{eqnarray}
where $V$ is the two-body interaction kernel and $G_{BS}$ is
the free two-body propagator.  In terms of the one body
propagator $G_i$ defined in Eq.~(\ref{def1}) below, the BS
propagator is
\begin{equation}
G_{BS}\equiv -i\,G_1 G_2\, .
\end{equation}
From Eq.~(\ref{2bodyeq02}) we have
\begin{equation}
V={\cal M}+{\cal M}G_{BS} V\, ,
\end{equation}
and substituting this equation into Eq.~(\ref{2bodyeq01}) gives the
following nonlinear equation for ${\cal M}$
\begin{equation}
{\cal M}=V-{\cal M}G_{BS} {\cal M}-{\cal M}G_{BS} V G_{BS} {\cal M}\, .
\label{2bodyeq03}
\end{equation}
\begin{figure}[t]
\begin{center}
\mbox{
   \epsfxsize=4.0in
   \epsfysize=1.0in
\epsfbox{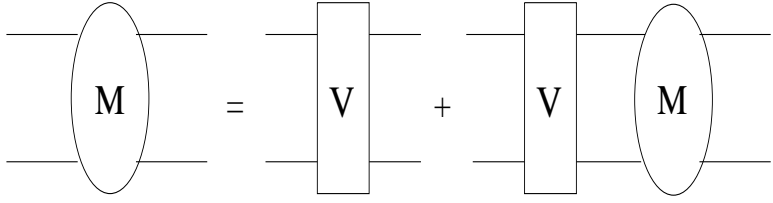}
%\epsffile{2_body_scatt_mat.fig.eps}
}
\end{center}
\caption{Diagrammatic representation of the two-body BS equation
for the scattering matrix.}
\label{2_body_scatt_mat}
\end{figure}

If the two-body system has a bound state at $P^2=M^2$, where $P$ is the
total four-momentum,  the scattering amplitude
$\cal M$ will have a pole at $P^2=M^2$.
Therefore, the two-body
scattering amplitude can be written
\begin{equation}
{\cal M}=-\frac{|\Gamma\rangle\,\langle{\Gamma} |}
{M^2-P^2}+{\cal R},
\label{2bodyeq04}
\end{equation}
where $|\Gamma\rangle$ is the Bethe-Salpeter vertex function
and $\langle{\Gamma} |$ is the representation of
$|\Gamma\rangle$ in the dual space, including Dirac
conjugation.  The first term is  the
contribution from the bound state propagator, with possible spin
degrees of freedom associated with the propagation of the bound
state suppressed, and $\cal R$ is a  finite at
$P^2=M^2$.  Substituting this form (\ref{2bodyeq04}) into the
linear  form (\ref{2bodyeq01}) of the equation for $\cal M$, and
equating residues at $P^2=M^2$ gives  the  wave equation for the
vertex function $|\Gamma\rangle$
\begin{equation}
|\Gamma\rangle=-VG_{BS}|\Gamma\rangle .
\label{2bodyeq045}
\end{equation}

Substituting (\ref{2bodyeq04}) into the nonlinear form of the
scattering equation  (\ref{2bodyeq03}), and keeping all terms
which are singular near the bound state pole at $P^2=M^2$, gives
the following relation
\begin{eqnarray}
|\Gamma\rangle\,\langle{\Gamma}|=&&\lim_{P^2\rightarrow
M^2}
\left\{
|\Gamma\rangle\;
\biggl[\frac{\langle{\Gamma}|\,G_{BS}
\left(1 +VG_{BS}\right) |\Gamma\rangle}{M^2-P^2}\biggr]
\;\langle{\Gamma}| \right.\nonumber\\
&&\left.  -  {\cal
R}\,G_{BS}\,(1+V\,G_{BS})\,|\Gamma\rangle\,
\langle{\Gamma}|
-|\Gamma\rangle\,\langle{\Gamma}|(1+G_{BS}\,V)\,
G_{BS}\,{\cal R}
\right\}\,
.\label{2bodyeq046}
\end{eqnarray}
Near the bound state pole at $P^2=M^2$ the BS equation
(\ref{2bodyeq045}) ensures that the last two terms involving the
remainder $\cal R$ vanish. Therefore, as
$P^2\rightarrow M^2$, Eq.~(\ref{2bodyeq046}) reduces to
\begin{equation}
1=\lim_{P^2\rightarrow
M^2}\frac{\langle{\Gamma}|\,G_{BS}\left( 1+
VG_{BS}\right)|\Gamma\rangle}{M^2-P^2}\, .
\end{equation}
Since both the numerator and the denominator vanish in this limit, we
expand the numerator around $P^2=M^2$, giving
\begin{equation}
1=-\langle{\Gamma}|\;\frac{\partial G_{BS}}{\partial
P^2}\;|\Gamma\rangle
-\langle{\Gamma}|\;\frac{\partial~}{\partial P^2}
\left( G_{BS}VG_{BS}\right) |\Gamma\rangle\, .
\label{2bodyeq05}
\end{equation}
(The BS equation ensures that any terms proportional to
$\partial|\Gamma\rangle/\partial P^2$ will also vanish.)  Distributing the
derivative in the second term over the product
$G_{BS}VG_{BS}$ and using the bound-state BS equation
(\ref{2bodyeq045}) reduces the normalization condition to
\begin{equation}
1=\langle{\Gamma}|\;G'_{BS}\;|\Gamma\rangle
-\langle{\Gamma}|\; G_{BS}V'G_{BS}\;|\Gamma\rangle\, .
\label{2bodyeq06}
\end{equation}
where
\begin{equation}
G'_{BS}=\frac{\partial G_{BS}}{\partial P^2}={P^\mu\over2P^2}
{\partial G_{BS}\over\partial P^\mu}
\, , \qquad
V'=\frac{\partial V}{\partial P^2} ={P^\mu\over2P^2}
{\partial V\over\partial P^\mu}\, ,
\end{equation}
and all derivatives are taken at the bound state pole
$P_0 = E(\vec{P})= (\vec{P}^{\, 2} + M^2 )^{1/2}$.

\begin{figure}[t]
\begin{center}
\mbox{
   \epsfxsize=5.0in
\epsfbox{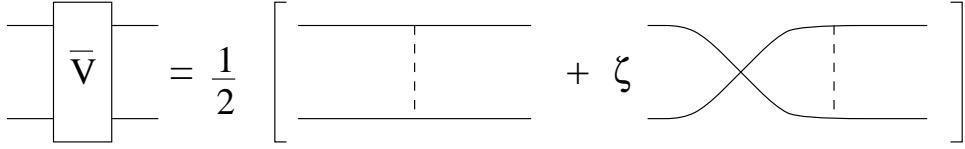}
%\epsffile{symmetrized.fig.eps}
}
\end{center}
\caption{Diagrammatic representation of the symmetrized kernel.}
\label{symmetrized}
\end{figure}

For identical particles, the symmetrized scattering matrix,
which we will denote by $M$, must satisfy
\begin{equation}
{\cal P}_{12}{M}={M}{\cal P}_{12}=\zeta{M} \label{symmxx}
\end{equation}
where ${\cal P}_{12}$ is the permutation operator that exchanges all labels
for  particles 1 and 2, and $\zeta=1$ for bosons and $\zeta=-1$ for
fermions. This can be achieved by symmetrizing the
unsymmetrized scattering matrix ${\cal M}$ according to
\begin{equation}
{M}={\cal A}_2 {\cal M}
= {\cal A}_2 {\cal M}{\cal A}_2 \, ,
\label{Maverage}
\end{equation}
where
\begin{equation}
{\cal A}_2= \frac{1}{2}\left( 1+\zeta{\cal P}_{12}\right)
\label{A2}
\end{equation}
is the symmetrization operator and
the last relation follows from ${\cal A}_{2}{\cal M}=
{\cal M}{\cal A}_{2}$ and ${\cal A}_2^2={\cal A}_2$.
Using the Bethe-Salpeter equation for the scattering
matrix yields
\begin{eqnarray}
{M}={\cal A}_2 {\cal M}&=& {\cal A}_2
\left(V-VG_{BS}{\cal M}\right)\nonumber\\
&=& {\cal A}_2 V-
{\cal A}_2 VG_{BS} {\cal A}_2{\cal M}\nonumber\\
&=&{\cal A}_2 V-
{\cal A}_2 VG_{BS} M
\end{eqnarray}
where the identities $ {\cal A}_2 V={\cal A}_2 V{\cal A}_2$ and
${\cal A}_2 G_{BS}=G_{BS}{\cal A}_2$ have been used in rewriting
the expression.  Defining the symmetrized kernel illustrated
diagrammatically in Fig.~\ref{symmetrized}
\begin{equation}
\overline{V}={\cal A}_2 V =\frac{1}{2}
\left( 1+\zeta{\cal P}_{12}\right) V \, ,
\label{Vaverage}
\end{equation}
gives a Bethe-Salpeter equation for the symmetrized
scattering matrix with the same form as the Bethe-Salpeter
equation for nonidentical particles
\begin{equation}
{M}=\overline{V}-\overline{V}G_{BS}{ M}\, .
\label{Gross2sym}
\end{equation}

%\input figure2a % symmetrized.fig.eps
%\begin{figure}[thb]

Since the structure of this equation is the same as in the case of identical
particles, the derivation of the bound state equation and the normalization
condition is the same as before with
replacement of $V\rightarrow\overline{V}$
\begin{equation}
1={_s}\!\langle{\Gamma}|\;G'_{BS}\;|\Gamma\rangle\!_s
-{_s}\!\langle{\Gamma}|\;
G_{BS}\overline{V}'G_{BS}\;|\Gamma\rangle\!_s\, ,
\label{2bodyeq06s}
\end{equation}
where
\begin{equation}
|\Gamma\rangle\!_s={\cal A}_2|\Gamma\rangle
\end{equation}
is the symmetrized vertex function.

Note that the symmetrized scattering matrix defined
above is normalized differently from the symmetrized
scattering matrices presented in most field theory texts, as
described in the Appendix.

\subsection{Normalization of the Gross two-body vertex functions}

\begin{figure}[t]
\begin{center}
\mbox{
   \epsfysize=1.0in
\epsfbox{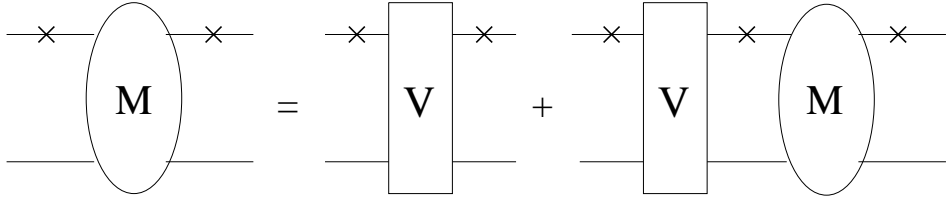}
%\epsffile{2_body_scatt_mat_gr.fig.eps}
}
\end{center}
\caption{Diagrammatic representation of the two-body Gross equation
for the scattering matrix.  The $\times$ on the line for particle 1 indicates
that it is on shell.}
\label{2_body_scatt_mat_gr}
\end{figure}

The Gross or spectator equation for distinguishable particles can be obtained
from the corresponding Bethe-Salpeter equation by means of a simple
prescription. For unequal mass particles with relatively long-range
interactions, the heavier of the two particles is placed on its
positive-energy mass shell in performing the energy loop integral over
intermediate state four-momenta. Here, the heavier particle is assumed to
particle 1. The scattering matrix then becomes
\begin{equation}
{\cal M}=V-V{\cal Q}_1{G}_2{\cal M} \label{GrossM}
\end{equation}
where the projection operator ${\cal Q}_i={\cal Q}_i^2$ places
particle $i$ on its positive-energy mass shell. Equation
(\ref{GrossM})  does  not represent a closed set of solvable
equations. In order to obtain these  it is necessary to also
place particle 1 on shell in both the initial and  final states
leading to
\begin{equation}
{\cal M}_{11}=V_{11}-V_{11}G_2{\cal M}_{11} \label{Gross11}
\end{equation}
where $V_{ii}={\cal Q}_i V{\cal Q}_i$ and ${\cal M}_{ii}={\cal Q}_i{\cal M}
{\cal Q}_i$.  This is illustrated in
Fig.~\ref{2_body_scatt_mat_gr}.

\begin{figure}[b]
\begin{center}
\mbox{
   \epsfysize=0.8in
\epsfbox{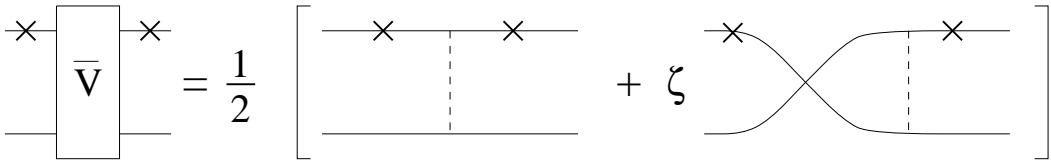}
%\epsffile{symmetrized_gr.fig.eps}
}
\end{center}
\caption{Diagrammatic representation of the symmetrized kernel for Gross
equation.  As in Fig.~3, the $\times$ indicates that the particle is on shell.}
\label{symmetrized2}
\end{figure}
 
The situation for identical particles is somewhat more complicated than for
the Bethe-Salpeter equation. Since ${\cal P}_{12}G_2{\cal P}_{12}=G_1$ and
${\cal P}_{12}{\cal Q}_1{\cal P}_{12}={\cal Q}_2$, the scattering matrix
can no longer be symmetrized by simply  symmetrizing the final state. The
solution to this problem is to use a  symmetric version of the spectator
equation
\begin{equation}
{\cal M}=V- \frac{1}{2}\,V\left({\cal Q}_1
G_2+{\cal Q}_2 G_1\right) {\cal M} \label{eqsyxx}
\end{equation}
where the factor of $\frac{1}{2}$ is necessary to prevent
double counting  of the elastic unitary cut of the scattering
matrix.  The factor of $\frac{1}{2}$ emerges automatically
if we think of obtaining Eq.~(\ref{eqsyxx}) from the BS
equation by doing the integration over the relative energy
variable by averaging the results obtained from closing the
contour in the upper half and lower half planes (and keeping
the positive energy nucleon poles only).   Eq.~(\ref{eqsyxx})
can now be  symmetrized in the same way as the
Bethe-Salpeter equation,  giving
\begin{eqnarray}
{M}&=&\overline{V}-\frac{1}{2}\,\overline{V}\left( {\cal Q}_1
G_2+{\cal Q}_2 G_1\right)
{\cal M}\nonumber\\
&=&\overline{V}-\frac{1}{2}\overline{V}\left({\cal Q}_1 G_2
+{\cal P}_{12}{\cal Q}_1  G_2{\cal P}_{12}\right)
 {\cal M}\nonumber\\
&=&\overline{V}-\frac{1}{2}\overline{V}{\cal Q}_1 G_2\left(
1+\zeta {\cal P}_{12}\right) {\cal M}\nonumber\\
&=&\overline{V}-\overline{V}{\cal Q}_1 G_2 { M}\, .
\label{MintAvg}
\end{eqnarray}
Clearly, ${\cal Q}_1 G_2$ could be eliminated in favor of
${\cal Q}_2 G_1$ in exactly the same manner. Therefore, the
$M$-matrix is independent of the choice of on-shell particle,
and the resulting integral equation for the scattering matrix
for (\ref{MintAvg}) can be taken to be
\begin{equation}
{M}_{11}=\overline{V}_{11}-\overline{V}_{11}G_2{M}_{11} \, ,
\end{equation}
where the symmetrized potential is shown in
Fig.~\ref{symmetrized2}.

\begin{figure}[t]
\begin{center}
\mbox{
   \epsfxsize=5in
\epsfbox{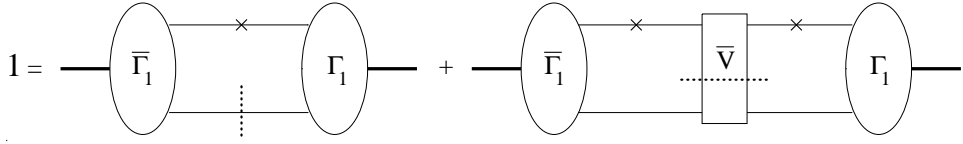}
%\epsffile{2_body_gross_norm.fig.eps}
}
\end{center}
\caption{Diagrammatic representation of the normalization condition for the
two-body Gross vertex function.  The dashed line represents the derivative
$\partial/\partial P^2$, and the $\times$ means that the particle is on shell.}
\label{2_body_gross_norm}
\end{figure}

As for the Bethe-Salpeter equation, the existence of a bound state implies
a pole in the scattering matrix,
\begin{equation}
{M}_{11}=-\frac{|\Gamma_1\rangle\,
\langle{\Gamma}_1|}
{M^2-P^2}+{\cal
R}\, ,
%\label{2bodyeq04}
\end{equation}
where $|\Gamma_1\rangle$ is the bound state vertex function
with particle 1 on  shell.  Repeating the development used in
the preceding section gives the Gross equation for
$|\Gamma_1\rangle$
\begin{equation}
|\Gamma_1\rangle=-\overline{V}_{11}G_{2}|\Gamma_1\rangle \, .
\label{2bodyeq045g}
\end{equation}
We will choose the momentum of the on-shell particle and the
total momentum $P$ as independent variables so that
$\partial/\partial P_\mu$ refers to differentiation with
respect to $P_\mu$ holding {\it the momentum of the on-shell
particle constant\/}.  Hence the on-shell projection
operators ${\cal Q}$ will be independent of $P_\mu$, and the
normalization condition becomes:
\begin{equation}
1=\langle{\Gamma}_1|\,G'_2\,|\Gamma_1\rangle
-\langle{\Gamma}_1|\, G_2 \,\overline{V}'_{11}\,G_2\,
|\Gamma_1\rangle\, .
\label{GrossNorm}
\end{equation}
This equation is represented diagrammatically in
Fig.~\ref{2_body_gross_norm}.  Note that this result could also
be obtained by applying the spectator prescription directly to
the unsymmetrized version of the Bethe-Salpeter normalization
condition (\ref{2bodyeq06s}).  Clearly, a similar expression can
be derived for $\Gamma_2$.

\subsection{Conservation of electromagnetic charge for the Gross
equation}

Since we have obtained the normalization condition
(\ref{GrossNorm}) directly from the two-body equation, with no
reference to the charge of the bound state, we now are in a
position to {\it prove\/} that the charge of the bound state is
conserved; ie. that the charge of the bound state is the sum of
the charges of its two constituents, regardless of its structure.

\begin{figure}[t]
\begin{center}
\mbox{
   \epsfxsize=6.0in
\epsfbox{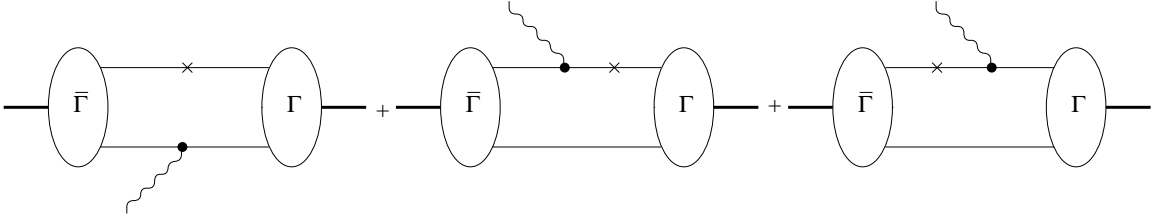}
%\epsffile{2_body_gross_formf.fig.eps}
}
\end{center}
\caption{Diagrams which contribute to the electromagnetic charge of the two
body bound state.}
\label{2_body_gross_formf}
\end{figure}

The construction of the matrix elements for the two-body current for the
Gross equation have been discussed by Gross and Riska\cite{GR} and
applied to elastic electron scattering from deuteron in \cite{deutletter}.
As argued in \cite{deutletter}, the bound state matrix elements of 
the electromagnetic current in the simple case when the OBE kernel
is constructed from the the exchange of {\it neutral\/} bosons is
represented by the Feynman diagrams shown in Fig.~\ref{2_body_gross_formf}.
Notice that since in our spectator formulation only the first particle is 
being placed on-shell, the first diagram in Fig.~\ref{2_body_gross_formf}
with $1 \leftrightarrow 2$ does not appear. Here we show explicitly that
the last two diagrams in Fig.~\ref{2_body_gross_formf} give a contribution
in the limit $q \to 0$  equal to that of the first diagram 
as well as the derivative term of the potential as it appears in the
normalization condition (\ref{GrossNorm}).  

Our argument will require that we replace the operator form of the
Gross equation (\ref{2bodyeq045g}) by its explicit representation
in momentum space.  We will use the notation
\begin{equation}
N=\left\{\begin{array}{cl}
1 & {\rm for\ bosons}\\
2m & {\rm for\ fermions}
\end{array}\right.
\end{equation}
and
\begin{equation}
{\cal Q}_i(\hat p)\to\Lambda_i(\hat p)=\left\{\begin{array}{cl}
1 & {\rm for\ bosons}\\
{\displaystyle {\not\!\hat p +m_i\over 2m_i}} & {\rm for\ fermions}
\end{array}\right. \qquad
G_i( p)=\left\{\begin{array}{cl}
{\displaystyle{\frac{\ \ 1}{m_i^2-p^2 - i \epsilon }}} & {\rm for\ bosons}\\
{\displaystyle{\frac{\ \ 1}{m_i-\not\!p - i \epsilon }}} & {\rm for\ fermions\, ,}
\end{array}\right. 
\label{def1}
\end{equation}
where $\Lambda_i$ is what remains of the
projection operator ${\cal Q}_i$ after the integral over the relative
energy has been carried out and the delta function which places particle 1
on shell has been removed,  and
$\hat p$ denotes an on-shell four-momentum with  $\hat
p^2=m^2$. The equation then becomes
\begin{equation}
\Gamma(\hat p_1;P)=-\int\frac{d^3 k_1\,N}{(2\pi)^3 \,2E_{k_1}}
\overline{V}(\hat p_1,\hat k_1;P)\Lambda_1(\hat k_1){G}_2(P-\hat
k_1) \Gamma(\hat k_1;P)\, .\label{GrossVertex}
\end{equation}
Note that the vertex  functions are written as a function of the
four-momentum of particle 1, $\hat p_1$,  and the total four-momentum,
$P$.  Multiplying both sides of the equation by the projection operator
$\Lambda_1(\hat p_1)$ and using $\Lambda_1^2(\hat k_1)=\Lambda_1(\hat
k_1)$ gives
\begin{equation}
\Gamma_1(\hat p_1;P)=-\int\frac{d^3 k_1\,N}{(2\pi)^3 \,2E_{k_1}}
\overline{V}_{11}(\hat p_1,\hat k_1;P){G}_2(P-\hat
k_1) \Gamma_1(\hat k_1;P)\, ,\label{GrossVertex2}
\end{equation}
where $\Lambda_1(\hat p_1) \overline{V}(\hat p_1,\hat k_1;P)
\Lambda_1(\hat k_1)=\overline{V}_{11}(\hat p_1,\hat k_1;P) $
and $\Gamma_1(\hat p_1;P)=\Lambda_1(\hat p_1) \Gamma(\hat p_1;P)$.
 
Using the Feynman rules, and the notation used in Eq.~(\ref{GrossVertex})
and (\ref{GrossVertex2}), the diagrams in Fig.~\ref{2_body_gross_formf}
yield 
\begin{eqnarray}
{\cal J}^\mu(P',P)&=& \int\frac{d^3 p_1\, N}{(2\pi)^3\, 2E_{p_1}}
\overline{\Gamma}_1(\hat p_1;P')\,{G}_2(P'-\hat p_1)
j^\mu_2(P'-\hat p_1,P-\hat p_1){G}_2(P-\hat p_1)\,\Gamma_1(\hat p_1;P)
\nonumber \\
&&\hspace*{-0.7cm} +\int\frac{d^3 p_1\,N}{(2\pi)^3\, 2E_{p_1}}
\overline{\Gamma}(\hat p_1+q;P'){G}_1(\hat p_1+q)
j^\mu_1(\hat p_1+q,\hat p_1){G}_2(P-\hat p_1)
\,\Gamma_1(\hat p_1;P) \nonumber \\
&&\hspace*{-0.7cm} +\int\frac{d^3 p'_1\,N}{(2\pi)^3\, 2E_{p'_1}}
\overline{\Gamma}_1(\hat p'_1;P')\,
j^\mu_1(\hat p'_1,\hat p'_1-q){ G}_1(\hat p'_1-q) {G}_2(P'-\hat p'_1)
\,\Gamma(\hat p'_1-q;P)   \, ,
\label{elasticME}
\end{eqnarray}
where $j^\mu_i(p^{\, \prime},p )$ is the renormalized e.m.\ current
of $i$-th constituent particle. When taken between the on-shell states
for $p^{\, \prime} = p$, this current defines a physical charge $e_i$.  
The one-body currents satisfy the  one-body Ward-Takahashi 
identity
\begin{equation}
q_\mu j^\mu_i(p'_i,p_i)= -e_i \left( {\tilde G}^{-1}_i(p')
-{\tilde G}^{-1}_i(p)\right)   
 \simeq  -e_i\,q_\mu \,\frac{\partial\ }{\partial p_\mu} 
 {\tilde G}^{-1}_i(p)\, ,
\label{WardTakahashi}
\end{equation}
where ${\tilde G}$ is a full one-particle propagator and  
the last form of the identity holds near the point 
$q_{\mu} \rightarrow 0$. Comparing the terms linear in 
$q_{\mu}$ we get from (\ref{WardTakahashi})
\begin{equation}
j^\mu_i(p,p)=  -e_i\,\frac{\partial\ }{\partial p_\mu}
{\tilde G}^{-1}_i(p) \,  .
\label{onebody0}
\end{equation}
Any purely transverse parts  $ j^\mu_{T,i}(p^{\, \prime},p )$ 
of the off-shell one-body current, which satisfy
$q_{\mu} j^\mu_{T,i}(p^{\, \prime},p ) =0$ and cannot be determined 
from (\ref{WardTakahashi}), are assumed to vanish in the limit
$q \to 0$. Therefore, they do not affect the charge $e_i$. 
Next, note the identity
\begin{equation}
{\tilde G}_i(p)\,j^\mu_i(p,p)\, {\tilde G}_i(p)=
 -e_i\,{\tilde G}_i(p)
\left(\frac{\partial\ }{\partial p_{\mu}}{\tilde G}^{-1}_i(p)\right)
{\tilde G}_i(p)=e_i\,\frac{\partial\ }{\partial p_{\mu}}
 {\tilde G}_i(p)  \, .
\label{onebody1}
\end{equation}
Thus, under the assumptions discussed above the insertion of the
$q = 0$ photon into the line of the constituent particle is 
unambigously determined by its physical charge and derivative of
its propagator. In this paper
we always approximate ${\tilde G}$ by (\ref{def1}) with a physical
mass $m_i$; our arguments are however valid also for the generalized
 ${\tilde G}_i(p)=h^2(p^2)/(m_i -\not\! p)$ used in a dynamical
model of Ref.~\cite{GVOH}.  
 
Now, the charge of the bound system is found by taking the limit of 
(\ref{elasticME})  $q\rightarrow 0$, or $P'\rightarrow P$, 
since contributions from any higher multipoles, such as magnetic dipole 
or charge quadrupole must vanish for $q =0$.  
In the limit considered the
propagators for particle 1 in the last two terms, ${G}_1(\hat
p_1+q)$ and ${G}_1(\hat p'_1-q)$, become singular.  However, it is easy
to see that these singularities cancel to give a finite result.

\begin{figure}[t]
\begin{center}
\mbox{
   \epsfxsize=4.0in
   \epsfysize=1.4in
\epsfbox{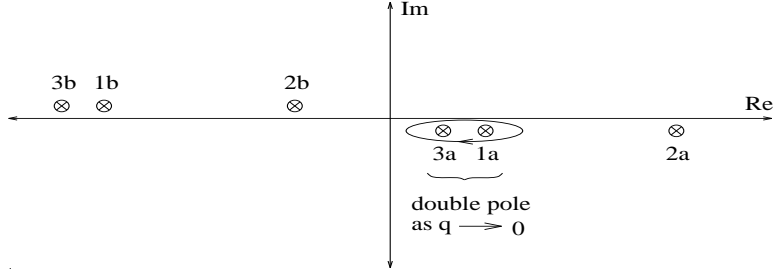}
%\epsffile{2_body_poles.fig.eps}
}
\end{center}
\caption{Location in the $p_{10}$ complex plane of the 6 poles from the three
nucleon propagators in Eq.~(2.34).  The last two terms in Eq.~(2.33) emerge if
the integral of $p_{10}$ over the contour $C$ enclosing the poles 3a and 1a is
evaluated using the residue theorem.}
\label{2_body.poles}
\end{figure}

Before we proceed with the evaluation of the $q\rightarrow 0$ limit of
Eq.~(\ref{elasticME}), it is instructive to describe this cancellation of
singularities in very general terms.  To this end we note that the {\it
last two terms\/} of Eq.~(\ref{elasticME}) can be identically represented
by the four dimensional integral
\begin{eqnarray}
{\cal J}^\mu(P',P)|_{{\hbox{\tiny last
2}}\atop{\hbox{\tiny terms}}}=-i\int_C\frac{d^4 p_1}{(2\pi)^4}&&
\overline{\Gamma}(p_1+q;P'){G}_1(p_1+q)
j^\mu_1(p_1+q,p_1)\nonumber\\
&&\quad\times G_1 (p_1) {G}_2(P- p_1)
\,\Gamma( p_1;P)\, ,
\label{contour}
\end{eqnarray}
where the integration over $p_{10}$ is done along the contour $C$ in 
the complex $p_{10}$ plane, shown in Fig.~\ref{2_body.poles}. 
This contour surrounds the positive energy poles
of the propagators ${G}_1(p_1+q)$ (at 3a) and ${G}_1(p_1)$ (at 1a), and
evaluating the integral by the residue theorem gives {\it only}\/ the last two
terms in Eq.~(\ref{elasticME}), where for notational simplicity in the
third term the integral over $d^3p_1$ is replaced by the integral over
$d^3p'_1$ (where ${\bf p}'_1={\bf p}_1+ {\bf q}$).  
It has been shown \cite{Gr65} that for elastic reactions 
the cuts of $\Gamma(p_1;P)$ never overlap with poles 3a and 1a
of Fig.~\ref{2_body.poles}, this makes the representation (\ref{contour}) 
possible and unambiguous for all values of momentum transfer $q$. 
It is now clear that as
$q\rightarrow 0$ the poles at 3a and 1a coalesce into a double pole, giving
a finite result.  Calculation of the residue of the double pole requires a
calculation of the derivative of the rest of the integrand, which also
explains the appearance of derivatives in the final result below.

%\input figure7 % 2_body_poles.fig.eps
% (equation numbers to go in by hand)
 
We now return to the reduction of Eq.~(\ref{elasticME}).  
Substituting (\ref{onebody1})  into the
first term of Eq.~(\ref{elasticME}) gives
\begin{equation}
{\cal J}^\mu(P,P)|_{{\hbox{\tiny first}}\atop
{\hbox{\tiny term}}}=
e_2\,
\int\frac{d^3 p_1\, N}{(2\pi)^3\, 2E_{p_1}}
\overline{\Gamma}_1(\hat p_1;P')\,\frac{\partial\ }{\partial
P_\mu}{G}_2(P-\hat p_1)\,\Gamma_1(\hat p_1;P) \, , \label{Eq1}
\end{equation}
where $\partial/\partial(P-\hat p_1)_\mu$ was replaced by
$\partial/\partial P_\mu$, which is possible only
because $\hat p_1$ is an on-shell four-momentum, so that
all four of its components are independent of $P$.  We
will return to this expression after we have reduced the
last two terms.
 
The last two (singular) terms in Eq.~(\ref{elasticME}) can be reduced by
expressing the vertex function with both particles
off-shell in terms of the vertex function with particle 1 on-shell by 
using the equation
\begin{equation}
\Gamma(p_1;P)=-\int\frac{d^3 k_1\,N}{(2\pi)^3 \,2E_{k_1}}
\overline{V}(p_1,\hat k_1;P){G}_2(P-\hat
k_1) \,\Gamma_1(\hat k_1;P)\, ,\label{GrossVertexoff}
\end{equation}
where the kernel $\overline{V}(p_1,\hat k_1;P)$ now connects an incoming 
channel with particle 1 on-shell to an outgoing channel with {\it both\/}
particles off-shell and is obtained from
$\overline{V}(\hat p_1,\hat k_1;P)$ by replacing
the four momentum $\hat p_1= (E_{p}, {\bf p})$  by the
off-shell four-momentum $(P_0- E_{P-p}, {\bf p})$.
[Eq.~(\ref{GrossVertexoff}) is obtained
from the original Eq.~(\ref{GrossM}) by going to the bound state
pole in the usual way.] Using this equation to
iterate each of the off-shell vertex functions once yields
\begin{eqnarray}
{\cal J}^\mu(P',P)|_{{\hbox{\tiny last 2}}\atop{\hbox{\tiny terms}}}=
\int\frac{d^3 p'_1\,N}{(2\pi)^3\,2E_{p'_1}}
\int\frac{d^3 p_1\,N}{(2\pi)^3\,2E_{p_1}}
\overline{\Gamma}_1\,(\hat p'_1;P')&&{G}_2(P'-\hat p'_1)
j^\mu_{\rm eff}(\hat p'_1,P';\hat p_1,P)\nonumber\\
&&\times{G}_2(P-\hat p_1)\,\Gamma_1(\hat p_1;P)\label{elasticME2}
\end{eqnarray}
where
\begin{eqnarray}
j^\mu_{\rm eff}(\hat p'_1,P';\hat p_1,P)=&&
\int\frac{d^3 k_1\,N}{(2\pi)^3\,2E_{k_1}}
 \Lambda_1(\hat p'_1) \biggl[\overline{V}(\hat p'_1,\hat k_1+q;P') 
 {G}_1(\hat k_1+q) \nonumber\\
&& \times j^\mu_1(\hat k_1+q,\hat k_1) \Lambda_1(\hat k_1)  
 {G}_2(P-\hat k_1)\overline{V}(\hat k_1,\hat p_1;P) + \nonumber\\
 && \overline{V}(\hat p'_1,\hat k_1;P'){G}_2(P'-\hat k_1)
 \Lambda_1(\hat k_1) j^\mu_1(\hat k_1,\hat k_1-q) \nonumber\\
&&\quad\times{G}_1(\hat k_1-q)
\overline{V}(\hat k_1-q,\hat p_1;P)\biggr] \Lambda_1(\hat p_1)   \, ,
\label{jeff}
\end{eqnarray}
Using the WT identity (\ref{WardTakahashi}) for finite $q$, and
recalling that  $G_1^{-1}(\hat k_1)  \Lambda_1(\hat k_1)  =0$, 
gives for the divergence of the effective current
\begin{eqnarray}
q_\mu j^\mu_{\rm eff}(\hat p'_1,P';\hat p_1,P)&=&
-e_1 \Lambda_1(\hat p'_1) \int\frac{d^3 k_1\,N}{(2\pi)^3\,2E_{k_1}} 
  \nonumber\\
& &\times\biggl[ \overline{V} (\hat p'_1,\hat k_1+q;P')
 {G}_2(P-\hat k_1) \Lambda_1(\hat k_1)  \overline{V} (\hat k_1,\hat p_1;P)
\nonumber \\
&& -\overline{V} (\hat p'_1,\hat k_1;P') \Lambda_1(\hat k_1) 
 {G}_2(P'-\hat k_1)  
\overline{V} (\hat k_1-q,\hat p_1;P)\biggr] \Lambda_1(\hat p_1) \, .
\label{qjeff}
\end{eqnarray}
Note that this step removes the propagators $G_1(\hat k_1+q)$
and $G_1(\hat k_1-q)$ which are singular in the $q\to0$ limit.
The purely transverse contributions to the effective current 
(\ref{jeff}) arise
only from the transverse parts of the one-body current and so,
by assumption, vanish as $q \to 0$. 
Therefore, we can use the divergence of the effective
current to extract its finite part for $P^{\, \prime} = P$. 
As before we expand (\ref{qjeff}) about the point
$q=0$ and equate terms linear in $q_{\mu}$ to obtain
\begin{eqnarray}
j^\mu_{\rm eff}(\hat p'_1,P;\hat p_1,P)&=&
-e_1\int\frac{d^3 k_1\,N}{(2\pi)^3\,2E_{k_1}} \,
\Lambda_1(\hat p'_1)\nonumber\\
& &\times\biggl[\left(\frac{\partial\
}{\partial \hat k_{1\mu}}
 \overline{V}(\hat p'_1,\hat k_1;P)\right)
{G}_2(P-\hat k_1)\Lambda_1(\hat k_1)\overline{V}(\hat k_1,\hat p_1;P)
\nonumber \\
&& -\overline{V}(\hat p'_1,\hat k_1;P)
\left(\frac{\partial\ }{\partial P_{\mu}}{G}_2(P-\hat k_1)\right)
\Lambda_1(\hat k_1)
\overline{V}(\hat k_1,\hat p_1;P)
\nonumber \\
&& + \overline{V}(\hat p'_1,\hat k_1;P)
{G}_2(P-\hat k_1) \Lambda_1(\hat k_1)
\left(\frac{\partial\ }{\partial \hat
k_{1\mu}}\overline{V}_{11}(\hat k_1,\hat p_1;P)
\right)\biggr]\, \Lambda_1(\hat p_1)\, .
\label{qjeff0}
\end{eqnarray}
Substituting this into Eq.~(\ref{elasticME2}), and using
$\Lambda_1^2=\Lambda_1$ and the spectator equation (\ref{GrossVertex2})
for the bound state gives
\begin{eqnarray}
{\cal J}^\mu(P,P)|_{{\hbox{\tiny last 2}}\atop{\hbox{\tiny terms}}}=&&e_1
\int\frac{d^3 p_1\,N}{(2\pi)^3\,2E_{p_1}}
\overline{\Gamma}_1(\hat p_1;P)
\left(\frac{\partial\ }{\partial P_{\mu}}{G}_2(P-\hat p_1)\right)
\Gamma_1(\hat p_1;P) \nonumber\\
&&+e_1\int\int\frac{d^3 p'_1d^3 p_1\,N^2}{(2\pi)^6\,4E_{p'_1}E_{p_1}}
\overline{\Gamma}_1(\hat p'_1;P){G}_2(P-\hat p'_1)
\left(\frac{\partial\ }{\partial \hat p_{1\mu}}
 \overline{V}(\hat p'_1,\hat p_1;P)\right.\nonumber\\
&&+\left.\frac{\partial\ }{\partial
\hat p'_{1\mu}}\overline{V} (\hat p'_1,\hat p_1;P)
\right) {G}_2(P-\hat p_1)
\Gamma_1(\hat p_1;P)\, . \label{currentME3})
\end{eqnarray}
Note that the first term is identical to Eq.~(\ref{Eq1}) if
$e_1\to e_2$.  To reduce the last two terms use the fact that
the symmetrized one boson exchange (OBE) potential has the form
\begin{equation}
\overline{V}(\hat p'_1,\hat p_1;P)= \frac{1}{2}\left( 1+\zeta {\cal
P}_{12}\right) {V}(\hat p'_1,\hat p_1;P)=\frac{1}{2}\Bigl[ V_d(\hat
p_1-\hat p'_1)+\zeta V_e(\hat p_1+\hat p'_1-P)\Bigr] \, ,
\end{equation}
where $V_d$ is the {\it direct\/} term and $V_e$ the {\it exchange\/} term
as defined in Ref.~\cite{GVOH}. Therefore only the derivatives of the
exchange term contribute to the charge operator, and
\begin{eqnarray}
\left(\frac{\partial\ }{\partial \hat p_{1\mu}}
+\frac{\partial\ }{\partial \hat p'_{1\mu}}\right)
\overline{V}(\hat p'_1,\hat p_1;P)&=&
\frac{1}{2}\zeta\left(\frac{\partial\ }{\partial \hat p_{1\mu}}
+\frac{\partial\ }{\partial \hat p'_{1\mu}}\right)
V_e(\hat p_1+\hat p'_1-P)\nonumber\\
&=&-\zeta\frac{\partial\ }{\partial P_{\mu}}
V_e(\hat p_1+\hat p'_1-P)=
-2\frac{\partial\ }{\partial P_{\mu}}
\overline{V}(\hat p'_1,\hat p_1;P) \, , \label{Eq2}
\end{eqnarray}
where we again used the fact that $\hat p_1$ and $\hat
p'_1$ are both independent of $P$.  Combining all of these
results together gives our final result  for the charge
operator of the bound state
\begin{eqnarray}
{\cal J}^\mu(P,P)&=& (e_1+e_2)\,
\int\frac{d^3 p_1\, N}{(2\pi)^3\,2E_{p_1}}
\overline{\Gamma}_1(\hat p_1;P) {\partial G_2(\hat p_1;P)\over
\partial P_\mu}
\Gamma_1(\hat p_1;P)
\nonumber \\
&-&2e_1\,\int\int\frac{d^3 p'_1 d^3
p_1\,N^2}{(2\pi)^6\,4E_{p'_1}E_{p_1}}
\overline{\Gamma}_1(\hat p'_1;P) G_2(\hat p'_1;P)
{\partial\overline{V}_{11} (\hat p'_1,\hat p_1;P)\over
\partial P_\mu} G_2(\hat p_1;P)
\Gamma_1(\hat p_1;P)\Biggr]
\nonumber \\
&=&e_B\, 2P^\mu \, .
\end{eqnarray}
where we used $\Lambda_1^2=\Lambda_1$ and $\Lambda_1(\hat p'_1)\,
[{\partial }\overline{V}(\hat p'_1,\hat p_1;P)/{\partial P_{\mu}}]\,
\Lambda_1(\hat p_1) = {\partial }\overline{V}_{11}(\hat p'_1,\hat
p_1;P)/{\partial P_{\mu}}$.
In the last step we also assumed, for definiteness, that
the bound state has spin zero and total charge $e_B$.  Multiplying the
equation by $P_\mu$ and using the identity
\begin{equation}
P_\mu\left({\partial\, {\cal O}\over \partial P_\mu}\right)=2P^2
\left({\partial\, {\cal O}\over \partial P^2}\right)\, ,
\end{equation}
gives the relation
\begin{eqnarray}
e_B&=& (e_1+e_2)\,
\int\frac{d^3 p_1\, N}{(2\pi)^3\,2E_{p_1}}
\overline{\Gamma}_1(\hat p_1;P)  G'_2(\hat p_1;P)
\Gamma_1(\hat p_1;P)
\nonumber \\
&-&2e_1\,\int\int\frac{d^3 p'_1 d^3
p_1\,N^2}{(2\pi)^6\,4E_{p'_1}E_{p_1}}
\overline{\Gamma}_1(\hat p'_1;P) G_2(\hat p'_1;P)
\overline{V}'_{11} (\hat p'_1,\hat p_1;P) G_2(\hat p_1;P)
\Gamma_1(\hat p_1;P)\Biggr]\, .
\end{eqnarray}

We now distinguish two cases.  If the two particles are not
identical, the OBE kernel does not include an exchange term and
the $\overline{V}'_{11}$ term vanishes [recall Eq.~(\ref{Eq2})].
In this case the normalization condition gives  $e_B=e_1+e_2$.
If the particles are identical, the $\overline{V}'_{11}$ term
will not vanish, but the charges will be equal, so $e_1+e_2=2e_1=e_B$
and we obtain the same result.  For either identical or
distinguishable particles, the {\it normalization
condition ensures the conservation of charge.\/}
 
This proof can be generalized to kernels which include multiple
meson exchange.  In this case the additional energy dependence contained in
the kernel will lead to new contributions to the $V'$ term in the
normalization condition, but these new terms will be reproduced exactly
by the $q\to0$ limit of additional Feynman diagrams containing
interaction currents also required by the multiple meson exchange kernel,
and the charge will still be  conserved.

The charge conservation and its relation to the normalization
condition can be conveniently discussed in terms of N-particle
Ward-Takahashi identities \cite{Bentz}. We will present these 
identities together with the general e.m.\ currents for two-
and three-body systems for the spectator formalism in a
forthcoming paper \cite{prep}. 

\section{Normalization of Relativistic Three-Body Vertex Functions}

In this final section we find the normalization condition for
relativistic three-body vertex functions. The derivation for the  
three-body Bethe-Salpeter equation is formally identical to that
given above for the two-body case. We briefly repeat it just to set
a stage for discussion of more complicated spectator case.  
In particular, we discuss in some detail the symmetrization necessary
for the identical particles and at the end we express our normalization
condition in terms of the symmetrized Faddeev subamplitudes.

\subsection{Normalization of three-body Bethe-Salpeter vertex
functions}

\subsubsection{Distinguishable Particles}

The three-body Bethe-Salpeter equation for distinguishable
particles can be  written
\begin{eqnarray}
{\cal T}=&&{\cal V}
-{\cal V}G^0_{BS} {\cal T} \label{faddeev1a} \\
=&&{\cal V}
- {\cal T}G^0_{BS}{\cal V}\, ,
\label{faddeev2a}
\end{eqnarray}
where $G^0_{BS}=-G_1G_2G_3$ is the three body BS propagator, and
\begin{equation}
{\cal V}=i\,\sum_{i}V^i {G}^{-1}_i =\sum_{i} {\cal
V}^i\label{Vsym}
\end{equation}
is the sum of
all products of two body potentials
$V^i$ multiplied by the inverse of the propagator $-iG_i$ of the
third,  non-interacting spectator, and we are assuming that there
are no relativistic three body forces (ie. diagrams which
are three-body irreducible).  In these equations and
those which follow we have adopted the odd-man-out or
spectator notation. If
$i$, $j$ and
$k$ represent some permutation of 1, 2 and 3, the two-body
potential
$V^i$ represents the interaction of particles $j$ and $k$.

\begin{figure}[t]
\begin{center}
\mbox{
   \epsfysize=2.5in
\epsfbox{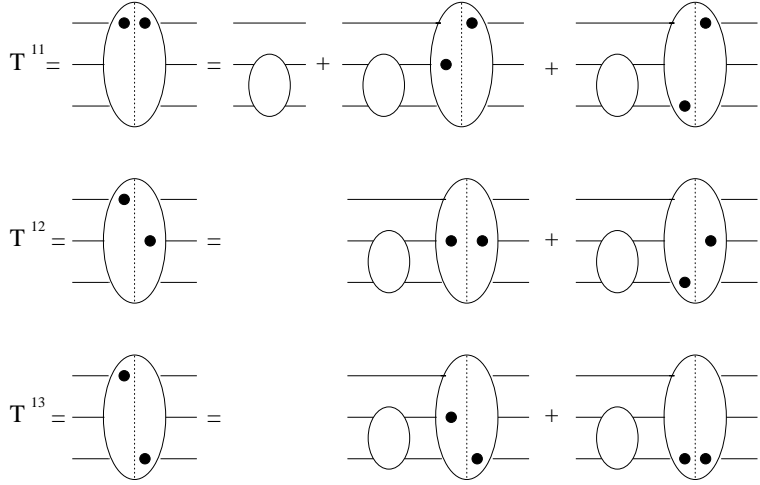}
%\epsffile{faddeev.fig.eps}
}
\end{center}
\caption{Diagrammatic representation of the Faddeev equations for the amplitudes
${\cal T}^{1i}$.  Note that the spectator is identified by the solid dot.}
\label{Faddeev}
\end{figure}

These equations look
deceptively simple, but the singularities in
${\cal V}$ make them difficult to solve, and they are usually reduced
to Faddeev form by introducing the subamplitudes
${\cal T}^{ii'}$ in which particle $i'$ is the initial spectator
and particle $i$ the final spectator. Since these objects
represent components of a summed Feynman  perturbation expansion
of the scattering problem, the terms ``initial''  and ``final''
refer to the topological form of the corresponding Feynman
diagrams, not to temporal priority.  Dropping the convention of
summation over repeated indices, so that {\it all sums will
be written explicitly\/}, the  initial equations for these
subamplitudes follow immediately from (\ref{faddeev1a}) and
(\ref{faddeev2a}):
\begin{eqnarray}
{\cal T}^{ii'}=&&i\,\delta_{ii'}V^i {G}^{-1}_i
-{V}^iG^i_{BS} \sum_{\ell} {\cal T}^{\ell i'} \label{faddeev1b}
\\ =&&i\,\delta_{ii'}V^i {G}^{-1}_i
-\sum_{\ell} {\cal T}^{i\ell}G^{i'}_{BS}V^{i'}\, ,
\label{faddeev2b}
\end{eqnarray}
where the singularities have been canceled using
$i\,G_i^{-1}G^0_{BS}=-i\,G_jG_k\equiv G^i_{BS}$.  Note that
\begin{equation}
{\cal V}^iG^0_{BS}=V^iG^i_{BS}\, . \label{defvi}
\end{equation}
We will
adopt the convention that the index $\ell$ can take on any
value, and that $i'$, $j'$, and $k'$ are all different indices
describing initial state particles (just as $i$, $j$, and $k$
describe final state particles). Note that
\begin{equation}
{\cal T}=\sum_{ii'} {\cal T}^{ii'}
\label{fadsumt}
\end{equation}
and that Eq.~(\ref{faddeev1a}) follows by summing (\ref{faddeev1b}) over $i$
and $i'$, and that Eq.~(\ref{faddeev2a}) follows by summing
(\ref{faddeev2b}) over $i$ and $i'$. Convergence of these
equations can be further improved by moving the
${\cal T}^{ii'}$ from the rhs to the lhs and introducing the
two-body scattering amplitude
\begin{equation}
{\cal M}^{i}=\left(1+{V}^iG^i_{BS}\right)^{-1}\,V^i=
V^i\,\left(1+{V}^iG^i_{BS}\right)^{-1}\, .
\label{matrix}
\end{equation}
This gives the famous Faddeev equations for the subamplitudes
\begin{eqnarray}
{\cal T}^{ii'}=&&i\,\delta_{ii'}{\cal M}^i {G}^{-1}_i
-{\cal M}^iG^i_{BS} \sum_{\ell\neq i} {\cal T}^{\ell i'}
\label{faddeev1}
\\ =&&i\,\delta_{ii'}{\cal M}^i {G}^{-1}_i
-\sum_{\ell\neq i'} {\cal T}^{i\ell}G^{i'}_{BS}{\cal M}^{i'}
\, .   \label{faddeev2}
\end{eqnarray}
The Feynman diagrams representing Eq.~(\ref{faddeev1}) are shown
in Fig.~\ref{Faddeev}.  In these figures the solid  dot denotes
the spectator.

Since the logic behind the resummation of the multiple scattering series in
terms of Faddeev amplitudes is the same as in the nonrelativistic three-body 
problem, it is not surprising that the equations have a form similar to
the nonrelativistic Faddeev equations \cite{Fad}. It is important to
remember, however, that in this case the summation is done in terms of
Feynman perturbative expansion rather than the time-ordered perturbation
theory of the nonrelativistic case. As a result global three-body propagators
are replaced by products of one-body Feynman propagators and an inverse
propagator must be included in the driving term to ensure that no propagator
appears for external legs of the three-body $T$ matrix. Similarly, due to
similarities in the underlying logic, many of the expressions derived
below have nonrelativistic analogues of similar form. The connection
between the relativistic formulation of the three-body problem and the
usual nonrelativistic approach will be explored elsewhere in detail.

A bound state of the three-body system is associated with the
occurrence of  a  pole at $P^2=M^2$ in the three-body scattering
amplitude ${\cal T}$ (and in each of the subamplitudes ${\cal
T}^{ii'}$),  where $P$ is the total momentum of the three-body
system and $M$ is the mass  of the bound state,  Therefore,
${\cal T}$ and ${\cal T}^{ii'}$ can be written as the sum of a
pole term and a  regular part $R^{ii'}$
\begin{eqnarray}
{\cal
T}^{ii'}=&&-\frac{|\Gamma^i\rangle\,
\langle{\Gamma}^{i'}\!|}
{M^2-P^2}+R^{ii'} \label{tij}\\ {\cal
T}=&&-\frac{|\Gamma\rangle\,\langle{\Gamma}|}
{M^2-P^2}+\sum_{ii'}R^{ii'} \, ,  \label{txx}
\end{eqnarray}
where $|\Gamma\rangle=\sum_i|\Gamma^i\rangle$.
Inserting the ansatz (\ref{tij}) into the scattering equations and taking the
limit $P^2\rightarrow M^2$ gives the equation for the bound state vertex
subamplitudes
\begin{equation}
|\Gamma^i\rangle=-{\cal M}^iG^i_{BS}
\sum_{\ell\neq i}|\Gamma^\ell\rangle\, .
\label{gammai}
\end{equation}
Similarly, using either Eq.~(\ref{matrix}) or starting with the
equations for ${\cal T}$, we obtain
\begin{equation}
|\Gamma\rangle=-\sum_{\ell}V^\ell
G^\ell_{BS}|\Gamma\rangle = -{\cal
V}G^0_{BS}|\Gamma\rangle \, .
\label{gammaix}
\end{equation}

The derivation of the three-body normalization condition is most easily
obtained from the ansatz (\ref{txx}) and the original equations
(\ref{faddeev1a}) and (\ref{faddeev2a}).  Repeating the steps which lead to
Eq.~(\ref{2bodyeq06}) we obtain for the three-body vertex function
the normalization condition of exactly the same form 
\begin{equation}
1=\langle{\Gamma}|\;\frac{\partial G^0_{BS}}{\partial
P^2}\; |\Gamma\rangle
-\langle{\Gamma}| G^0_{BS}\;\frac{\partial~{\cal
V}}{\partial P^2}\; G^0_{BS} |\Gamma\rangle =
-\langle{\Gamma}| G^0_{BS}\;\frac{\partial~}
{\partial P^2}\left({\cal V}\,G^0_{BS}\right)
|\Gamma\rangle \, .
\label{3bodyeq06}
\end{equation}
The two forms of the normalization condition (\ref{3bodyeq06}) 
are identical because of the bound state Eq.~(\ref{gammaix}).  

\subsubsection{Identical particles}

The equations and normalization condition for identical particles
can be obtained immediately by symmetrizing the results given
above.  However, to lay the foundation for the discussion of the
spectator equations it is convenient to describe this
symmetrization in detail. Most of the operator algebra below
is exactly the same as in non-relativistic case.

The normalized three-body antisymmetrization operator is
\begin{eqnarray}
{\cal A}_3\equiv&& \frac{1}{3!}\left( 1+\zeta{\cal
P}_{12}+\zeta{\cal P}_{13}+
\zeta{\cal P}_{23}+{\cal P}_{4}+{\cal P}_{5}\right)\nonumber\\
=&& \frac{1}{3!}\left( 1+\zeta{\cal P}_{ij}\right)\left(1+\zeta{\cal P}_{ik}+
\zeta{\cal P}_{jk}\right)\nonumber\\
=&& \frac{1}{3!}\left(1+\zeta{\cal P}_{ik}+
\zeta{\cal P}_{jk}\right) \left( 1+\zeta{\cal P}_{ij}\right)\, ,
\label{symmetry3}
\end{eqnarray}
where
\begin{eqnarray}
{\cal P}_{4}=&&{\cal P}_{13}{\cal P}_{12}={\cal P}_{23}{\cal P}_{13}
={\cal P}_{12}{\cal P}_{23}\nonumber\\
{\cal P}_{5}=&&{\cal P}_{23}{\cal P}_{12}={\cal P}_{12}{\cal P}_{13}
={\cal P}_{13}{\cal P}_{23} \, .
\end{eqnarray}
Note that ${\cal A}_3={\cal A}_3{\cal A}_2={\cal A}_2{\cal
A}_3$ where we may choose to have the two body symmetrization
operator ${\cal A}_2$ act on {\it any\/} of the three two-body
subspaces.  We will exploit this property in the following
discussion.

Using the relations ${\cal P}_{jk} V^i {\cal P}_{jk}=V^i$ and
${\cal P}_{ji}V^i {\cal P}_{ij}=V^j$ (which also hold for $G_i$)
we can readily show that the symmetrization operator
${\cal A}_3$ commutes with the total potential (\ref{Vsym}).
The symmetrized version of Eq.~(\ref{faddeev1a}) is then
obtained by multiplying the equation by ${\cal A}_3$, which
gives
\begin{eqnarray}
T={\cal A}_3 {\cal T}=&&{\cal A}_3{\cal V}
-{\cal A}_3{\cal V}G^0_{BS} {\cal T} \nonumber\\
=&&{\cal A}_3{\cal V}
- {\cal A}_3{\cal V}G^0_{BS} {\cal A}_3{\cal T} \nonumber\\
=&&\overline{\cal V}
- \overline{\cal V}G^0_{BS} T \, ,
\label{3bodysym}
\end{eqnarray}
where
\begin{eqnarray}
\overline{\cal V}=&& {\cal A}_3{\cal V}={\cal
A}_3\sum_i \overline{\cal V}^{\,i}\nonumber\\
\overline{\cal V}^{\,i}=&&i\,
\overline{V}^{\,i} {G}^{-1}_i \, .\label{Vsym2}
\end{eqnarray}
A similar argument works for the transposed equation
(\ref{faddeev2a}), and hence the symmetrized amplitude
$T$ satisfies the equations satisfied by ${\cal T}$ with
${\cal V}$ replaced by  $\overline{\cal V}$.
Note that we have written the kernel (\ref{Vsym2}) in terms of
the symmetrized two-body kernels (\ref{Vaverage}), and that
$\overline{\cal V}$ can also be written in terms of a single
$\overline{\cal V}^{\,i}$
\begin{equation}
\overline{\cal V}={\cal A}_3\sum_{\ell}
\overline{\cal V}^\ell = {\cal A}_3\sum_{\ell}{\cal P}_{\ell
i} \overline{\cal V}^{\,i} {\cal P}_{i\ell}= {\cal
A}_3\overline{\cal V}^{\,i}  \sum_{\ell} \zeta{\cal P}_{i \ell}
\, . \label{defvi2}
\end{equation}
As before, $\ell$ can be any
index, and consistency requires the definition ${\cal
P}_{ii}=\zeta$, so that $\zeta {\cal P}_{ii}=1$.  The last
step in (\ref{defvi2}) follows from ${\cal P}_{ji}
\overline{\cal V}^{\,i} {\cal P}_{ij}= \overline{\cal V}^{\,j}$
and ${\cal A}_3{\cal P}_{i\ell}=\zeta {\cal A}_3$.  The
symmetric version of Eq.~(\ref{defvi})
\begin{equation}
\overline{\cal V}^{\,i}G^0_{BS}=\overline{V}^{\,i}G^i_{BS}\, .
\label{defvix}
\end{equation}
will be useful in the following discussion.

The symmetrized bound state vertex function $|\Gamma\rangle\!_s=
{\cal A}_3|\Gamma\rangle={\cal A}_3\sum_i|\Gamma^i\rangle$
satisfies the equations
\begin{eqnarray}
|\Gamma\rangle\!_s=&&-\overline{\cal V}
G^0_{BS}|\Gamma\rangle\!_s  =-3{\cal
A}_3\overline{V}^{\,i}G^i_{BS}|\Gamma\rangle\!_s
\nonumber\\
{_s}\!\langle\Gamma| =&&-{_s}\!\langle\Gamma |
G^0_{BS}\overline  {\cal V} =-3\;{_s}\!\langle\Gamma|
G^i_{BS} \overline{V}^{\,i} {\cal A}_3 \, ,
\end{eqnarray}
where we used
\begin{eqnarray}
{\cal P}_{ij}|\Gamma\rangle\!_s=&& \zeta|\Gamma\rangle\!_s
\nonumber\\
\sum_\ell\zeta{\cal P}_{i\ell}|\Gamma\rangle\!_s=&&
3|\Gamma\rangle\!_s \, .
\end{eqnarray}
The normalization condition for this vertex function follows
readily from the previous derivation
\begin{equation}
1={_s}\!\langle{\Gamma}|\;\frac{\partial G^0_{BS}}
{\partial P^2}\;|\Gamma\rangle\!_s
-{_s}\!\langle{\Gamma}|
G^0_{BS}\;\frac{\partial~\overline{\cal V}} {\partial P^2}\;
G^0_{BS} |\Gamma\rangle\!_s =
-{_s}\!\langle{\Gamma}|
G^0_{BS}\;\frac{\partial~} {\partial P^2}
\left(\overline{\cal V}\,G^0_{BS}\right)
|\Gamma\rangle\!_s \, .
\label{3bodyeq07}
\end{equation}
Note that this equation has a structure identical to all of the
previous results.

To make a better comparison with the spectator equations, it is
convenient to obtain the BS equations and normalization
condition for the symmetrized subamplitudes.  First introduce
the operator
\begin{equation}
{\cal A}_{ii'}={1\over3!}\left(1+\zeta{\cal P}_{jk}\right)\,\zeta {\cal P}_{ii'}
= {1\over3!}\,\zeta {\cal P}_{ii'} \left(1+\zeta{\cal P}_{j'k'}\right) \, .
\label{aij}
\end{equation}
The two forms of ${\cal A}_{ii'}$ are identical regardless of the specific
values
of $i$ and $i'$.  If $i=i'$, then
${\cal P}_{ii'}=\zeta$ and ${\cal P}_{jk}={\cal P}_{j'k'}$ because both sets of
indices $\{ijk\}$ and $\{i'j'k'\}$ must assume the values 1, 2, and 3.  If
$i\ne i'$, then $i'$ must equal either $j$ or $k$.  For definiteness assume
$i'=j$.  Then we may use the identity
\begin{equation}
{\cal P}_{jk}{\cal P}_{ij}={\cal P}_{ij}{\cal P}_{ik} 
\label{Pindentity}
\end{equation}
to prove the equivalence.  An identical argument works if $i'=k$.  
These identities will be used frequently in the discussion which follows. 
Note that
\begin{equation}
\sum_i{\cal A}_{ii'}= {\cal A}_3=\sum_{i'}{\cal A}_{ii'} \, .
\label{aij2}
\end{equation}

Now we define the symmetrized subamplitudes $T^{ii'}$ and
$|\Gamma^i\rangle\!_s$
\begin{eqnarray}
{T}^{ii'}=&&\sum_{\ell\ell'} {\cal A}_{i\ell} \,
{\cal T}^{\ell\ell'} \, {\cal A}_{\ell' i'} \label{sym1}\\
|\Gamma^i\rangle\!_s=&&\sum_\ell {\cal A}_{i\ell} \,
|\Gamma^\ell\rangle
\,  . \label{sym1a}
\end{eqnarray}
Using (\ref{aij2}) gives the relations
\begin{eqnarray}
\sum_i T^{ii'}=&& {\cal A}_3 \sum_{\ell\ell'} {\cal T}^{\ell\ell'} \,
{\cal A}_{\ell' i'}\label{sumt1}\\
\sum_{ii'}T^{ii'}=&& {\cal A}_3 \sum_{\ell\ell'} {\cal T}^{\ell\ell'}
{\cal A}_{3}= {\cal A}_3{\cal T} {\cal
A}_{3} = {T}\label{sumt2} \\
\sum_i|\Gamma^i\rangle\!_s=&& {\cal A}_3 \sum_\ell
|\Gamma^\ell\rangle = {\cal A}_3|\Gamma\rangle=
|\Gamma \rangle\!_s\, . \label{sumt3}
\end{eqnarray}
From the definitions it follows immediately that
\begin{eqnarray}
{\cal P}_{jk} {T}^{ii'}=&&\zeta\,{T}^{ii'}={T}^{ii'} {\cal P}_{j'k'}
\label{sym2}\\
{\cal P}_{jk}\,|\Gamma^i\rangle\!_s
=&&\zeta\,|\Gamma^i\rangle\!_s\, .
\label{sym2a}
\end{eqnarray}
It also follows from (\ref{aij}) that
\begin{eqnarray}
{\cal P}_{ij} {T}^{ii'}=&&\zeta\,{T}^{ji'} \nonumber\\
{T}^{ii'} {\cal P}_{i'j'} =&& \zeta\,{T}^{ij'} \nonumber\\
{\cal P}_{ij}
\,|\Gamma^i\rangle\!_s=&&\zeta\,|\Gamma^j\rangle\!_s\, .
\label{sym.fad.3}
\end{eqnarray}
To prove these relations define, for some operators
${\cal O}^{\, 1},{\cal O}^{\, 2},{\cal O}^{\, 3}$,  
the operator $\tilde{\cal O}^{\,i}$
\begin{equation}
\tilde{\cal O}^{\,i}=\sum_{\ell} {\cal A}_{i\ell} \,
{\cal O}^{\,\ell}  \, .
\label{op1}
\end{equation}
Then
\begin{eqnarray}
{\cal P}_{ij} \tilde{\cal O}^{\,i}=&& {1\over6} \, {\cal P}_{ij}
\left(1+\zeta{\cal P}_{jk}\right)\,\sum_\ell\zeta {\cal P}_{i\ell}
{\cal O}^\ell
\nonumber\\
=&&{1\over6} \, \left(1+\zeta{\cal P}_{ik}\right)
{\cal P}_{ij}\,\left({\cal O}^i+ \zeta {\cal P}_{ij}
{\cal O}^j + \zeta {\cal P}_{ik} {\cal O}^k \right)
\nonumber\\
=&&{1\over6} \, \left(1+\zeta{\cal P}_{ik}\right)
\,\left({\cal P}_{ij}{\cal O}^i+ \zeta
{\cal O}^j + \zeta {\cal P}_{ik}{\cal P}_{jk} {\cal O}^k \right)
\nonumber\\
=&&\zeta\, \tilde{\cal O}^{\,j} \, .
\label{op2}
\end{eqnarray}
In the last step we used $\left(1+\zeta{\cal P}_{ik}\right) {\cal P}_{ik}=
\zeta\,\left(1+\zeta{\cal P}_{ik}\right)$.  The first of the relations
(\ref{sym.fad.3}) follows immediately from (\ref{op2}), and the second by
taking the hermitian conjugate.

Next, note that
\begin{equation}
{\cal A}_{i\ell}{\cal V}^\ell = {\cal
A}_{i\ell}\overline{\cal V}^{\,\ell} =
\overline{\cal V}^{\,i}{\cal A}_{i\ell} \, , \label{av1}
\end{equation}
and
\begin{eqnarray}
\sum_{\ell\ell'}{\cal A}_{i\ell}\,\delta_{\ell\ell'}{\cal
V}^\ell{\cal A}_{\ell'i'}=&&{1\over9}\sum_\ell
{\cal P}_{i\ell}\overline{\cal V}^{\,\ell} {\cal
P}_{\ell i'}
\nonumber\\ =&&{1\over9}\overline{\cal V}^{\,i}
\sum_\ell{\cal P}_{i\ell}{\cal P}_{\ell i'}
=\zeta\,{1\over3}\overline{\cal V}^{\,i} {\cal P}_{ii'}
\, ,
\label{av2}
\end{eqnarray}
where in the last step we have used
\begin{equation}
   \frac{1}{2} (1 + \zeta P_{jk}) 
   \sum_\ell{\cal P}_{i\ell}{\cal P}_{\ell i'} =
    \frac{1}{2} (1 + \zeta P_{jk}) 3 \zeta P_{i i'} \, .
\end{equation}
Starting from Eqs.~(\ref{faddeev1b}) or (\ref{faddeev2b}), using the
definition (\ref{sym1}) and the properties (\ref{av2}), (\ref{av1}),
(\ref{sumt1}), and (\ref{aij2}) gives the following equations for the
symmetrized subamplitudes
\begin{eqnarray}
{T}^{ii'}=&&i\,\zeta\,{1\over3}\overline{V}^{\,i} {G}^{-1}_i
{\cal P}_{ii'} -\overline{V}^{\,i}G^i_{BS} \sum_\ell {T}^{\ell i'}
\label{faddeev1symm} \\
=&&i\,\zeta\,{1\over3}{\cal P}_{ii'}\overline{V}^{\,i'} {G}^{-1}_{i'}
 - \sum_\ell {T}^{i\ell}G^{i'}_{BS} \overline{V}^{\,i'}\, .
\label{faddeev2symm}
\end{eqnarray}
Summing Eq.~(\ref{faddeev1symm}) over $i$ and $i'$ recovers
Eq.~(\ref{3bodysym}), demonstrating consistency.  Note that
Eq.~(\ref{faddeev1symm}) is equivalent to
\begin{eqnarray}
{T}^{ii}=&&i\,{1\over3}\overline{V}^{\,i} {G}^{-1}_i
 -\overline{V}^{\,i}G^i_{BS} \sum_\ell {T}^{\ell i}
\label{faddeev4symm} \\
{T}^{ii'}=&& \zeta\,{T}^{ii} {\cal P}_{ii'} \label{faddeev5symm}\, ,
\end{eqnarray}
where (\ref{faddeev5symm}) merely recovers the symmetry
(\ref{sym.fad.3}).  Using this symmetry again, we may  reduce
(\ref{faddeev4symm}) to an equation for the single amplitude $T^{ii}$
\begin{eqnarray}
{T}^{ii}=&&i\,{1\over3}\overline{V}^{\,i} {G}^{-1}_i  -
\overline{V}^{\,i} G^i_{BS} \left(1+\zeta{\cal P}_{ij}+
\zeta{\cal P}_{ik}
\right){T}^{ii}\nonumber\\
=&&i\,{1\over3}\overline{V}^{\,i} {G}^{-1}_i
- \overline{V}^{\,i} G^i_{BS} \left(1+\zeta{\cal P}_{ij}+
\zeta{\cal P}_{jk}{\cal P}_{ij}{\cal P}_{jk}
\right){T}^{ii}\nonumber\\
=&&i\,{1\over3}\overline{V}^{\,i} {G}^{-1}_i
- \overline{V}^{\,i} G^i_{BS}\left(1+2\zeta{\cal P}_{ij}\right){T}^{ii}
\, .
\label{faddeev3symm}
\end{eqnarray}
Hence it is sufficient to solve one equation for $i=i'$, and obtain
all other amplitudes from (\ref{faddeev5symm}).
Using Eq.~(\ref{Gross2sym}) for the symmetrized two-body
scattering amplitude, the equation for $T^{ii}$ may be further
reduced to
\begin{equation}
{T}^{ii}= i\,{1\over3} M^{i} {G}^{-1}_i
- 2\zeta\,M^i G^i_{BS}{\cal P}_{ij}{T}^{ii} \, . \label{sym4}
\end{equation}
Similarly, the equations for the symmetrized subvertex functions are
\begin{eqnarray}
|\Gamma^i\rangle\!_s=&&  -
\overline{V}^{\,i} G^i_{BS} \left(1+2\zeta{\cal P}_{ij}\right)
|\Gamma^i\rangle\!_s
 = - \overline{V}^{\,i} G^i_{BS} |\Gamma\rangle\!_s
\label{faddeev3symma}\\
=&& - 2\zeta\,M^i G^i_{BS}{\cal P}_{ij}|\Gamma^i\rangle\!_s \, .
\label{sym4a}
\end{eqnarray}

Using the definitions of the symmetric subvertex functions,
the normalization condition (\ref{3bodyeq07}) can be further simplified.
First note that
\begin{eqnarray}
&&\frac{\partial}{\partial P^2}\left(\overline{\cal V}
G^0_{BS}\right)={\cal A}_3
\frac{\partial}{\partial P^2}\left(\overline{V}^{\,i}
G^{i}_{BS}\right) \sum_{\ell} \zeta{\cal P}_{i \ell}=
{\cal A}_3 \left\{
\frac{\partial \overline{V}^{\,i} }{\partial P^2} G^{i}_{BS}
+ \overline{V}^{\,i}\frac{\partial G^{i}_{BS} }{\partial P^2}
 \right\} \sum_{\ell} \zeta{\cal P}_{i \ell}
\,  .
\end{eqnarray}
To obtain this we used Eqs.~(\ref{defvi2},\ref{defvix}) and 
the fact that the exchange operators ${\cal P}_{ij}$ are independent 
of $P^2$, so that
$\partial {\cal P}_{ij}/\partial P^2=0$  for any exchange
operator.  Substituting this relation into (\ref{3bodyeq07})
and using the bound state equation (\ref{faddeev3symma}) 
and ${\cal A}_3 |\Gamma\rangle\!_s= |\Gamma\rangle\!_s$ gives 
immediately
\begin{equation}
1=-3i\;{_s}\!\langle{\Gamma}^{i}|\,G_i
\;\frac{\partial~G^i_{BS}}{\partial
P^2}\,|\Gamma\rangle\!_s +3i\;
{_s}\!\langle{\Gamma}|\,
G^i_{BS}G_i\;\frac{\partial~\overline{V}^{\,i}}  {\partial P^2}
G^i_{BS}\;|\Gamma\rangle\!_s\, .
\label{3bodyeq08a}
\end{equation}

This condition can be expressed in terms of a single subvertex
function, which we will choose to be $|\Gamma^i\rangle\!_s$.
Denoting the operator $G^i_{BS}G_i\;
(\partial~\overline{V}^{\,i}/\partial P^2)\; G^i_{BS}$ by
${\cal O}_2^i$,  and remembering that this operator contains a factor
of $1+ \zeta{\cal P}_{jk}$, the second term reduces to
\begin{eqnarray}
{_s}\!\langle{\Gamma}|\,
{\cal O}_2^i\;|\Gamma\rangle\!_s =&&
{_s}\!\langle{\Gamma}^i|\left(
1+\zeta{\cal P}_{ij}+\zeta{\cal P}_{ik}\right)
{\cal O}_2^i \,\left(1+\zeta{\cal P}_{ij}+\zeta{\cal P}_{ik}\right)
|\Gamma^i\rangle\!_s\nonumber\\
=&&{_s}\!\langle{\Gamma}^i|\left(
1+2\zeta{\cal P}_{ij}\right)\,
{\cal O}_2^i\,
\left(1+2\zeta{\cal P}_{ij}\right)
|\Gamma^i\rangle\!_s
\end{eqnarray}
where we used ${\cal P}_{jk}{\cal P}_{ij}{\cal P}_{jk}={\cal
P}_{ik}\to {\cal P}_{ij}$ as the operators ${\cal P}_{jk}$ are
eliminated using (\ref{sym2a}) and ${\cal P}_{jk}^2=1$.
Similarly, since the operator ${\cal P}_{jk}$
commutes with
${\cal O}_1^i=G_i({\partial~G^i_{BS}}/{\partial P^2})$, the
first term becomes
\begin{eqnarray}
{_s}\!\langle{\Gamma}^{i}|
\;{\cal O}_1^i\,|\Gamma\rangle\!_s
=&&{_s}\!\langle{\Gamma}^{i}|{\cal O}_1^i \left(
1+\zeta{\cal P}_{ij} + \zeta{\cal P}_{ik}\right)
|\Gamma^i\rangle\!_s \nonumber\\
=&&{_s}\!\langle{\Gamma}^i|{\cal O}_1^i
\left(1+ 2\zeta{\cal P}_{ij}\right) |\Gamma^i\rangle\!_s  \, .
\end{eqnarray}
Combining these terms gives a normalization condition expressed
in terms of $|\Gamma^i\rangle\!_s$
\begin{eqnarray}
1=&&-3i\,{_s}\!\langle{\Gamma}^{i}|
\;G_i\frac{\partial~G^i_{BS}}{\partial
P^2}\,\left(1+2\zeta{\cal P}_{ij}\right)|\Gamma^i\rangle\!_s
\nonumber\\
&&+3i
{_s}\!\langle{\Gamma}^i|\left(1+2\zeta{\cal P}_{ij}\right)\,
G^i_{BS}G_i\;\frac{\partial~\overline{V}^{\,i}}  {\partial P^2}
G^i_{BS}\;\left(1+2\zeta{\cal P}_{ij}\right)
|\Gamma^i\rangle\!_s\, .
\label{3bodyeq08b}
\end{eqnarray}

We are now ready to turn our discussion to the three-body
spectator equation.

\subsection{Normalization of the Three-body Gross Vertex
Functions}

\begin{figure}[t]
\begin{center}
\mbox{
   \epsfxsize=5.5in
\epsfbox{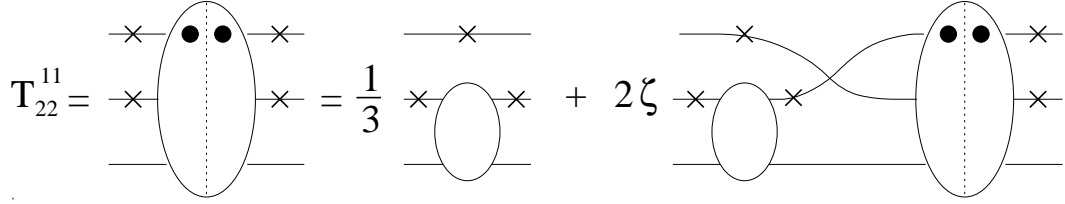}
%\epsffile{gross.fig.eps}
}
\end{center}
\caption{Diagrammatic representation of the Gross equation for the amplitude
$T^{11}_{22}$.  On-shell particles are labeled with an $\times$.  The final
state particles emerge from the left side of each diagram, and putting the final
state spectator on-shell consistently  automatically forces a second particle in
the final state to be on-shell.  In this way two particles are always on-shell.}
\label{Gross}
\end{figure}

The three-body spectator or Gross equation for identical
particles is  obtained from the symmetrized three-body
Bethe-Salpeter equation (\ref{sym4}) by placing the  spectator
particle on its positive energy mass shell \cite{Gross3b,SG}.  
In order to obtain a closed set of coupled equations a
second  particle must be placed on its mass shell in the
initial and final states, as illustrated in Fig. \ref{Gross}. 
Therefore, we  label those particles that are on-shell by replacing
subamplitudes
${T}^{ii'}$ by ${T}^{ii'}\rightarrow {T}^{ii'}_{jj'}=
{\cal Q}_i{\cal Q}_j{T}^{ii'}{\cal Q}_{i'}{\cal Q}_{j'}$.
Noting that ${\cal P}_{jj'}{\cal Q}_{j}{\cal P}_{jj'}=
{\cal Q}_{j'}$, and using Eqs.~(\ref{sym2}) and
(\ref{sym.fad.3}) we have
\begin{eqnarray}
{\cal P}_{ij}{T}^{ii'}_{jj'}=&& {\cal P}_{ij}{\cal Q}_i
{\cal Q}_j {T}^{ii'}{\cal Q}_{i'}{\cal Q}_{j'} =
{\cal Q}_j{\cal Q}_i\zeta{T}^{ji'}{\cal Q}_{i'}{\cal Q}_{j'}
=\zeta{T}^{ji'}_{ij'}\nonumber\\
{\cal P}_{ik}{T}^{ii'}_{jj'}=&& {\cal Q}_k
{\cal Q}_j {\cal P}_{ik}{T}^{ii'}{\cal Q}_{i'}{\cal Q}_{j'} =
\zeta {T}^{ki'}_{jj'}\nonumber\\
{\cal P}_{jk}{T}^{ii'}_{jj'}=&& {\cal Q}_i
{\cal Q}_k {\cal P}_{jk}{T}^{ii'}{\cal Q}_{i'}{\cal Q}_{j'} =
\zeta {T}^{ii'}_{kj'} \, .
\end{eqnarray}

Similar relations exist for the initial state.  Hence any
amplitude can be obtained by the action of the permutation
operators on the canonical amplitude ${T}^{ii}_{jj}$, and
it is necessary only to find and solve the equation for this
one amplitude. Note that we can choose $i$ and $j$ arbitrarily 
as long as $i \neq j$. The subamplitudes ${T}^{ii'}_{jj'}$ cannot
be added unless the same pairs of particles are on shell in both
the initial and final states. Therefore, there is no total
scattering amplitude such as (\ref{fadsumt}) 
for the 3-body spectator equation. 
The underlying physical justification for the use of the spectator 
equations is described in detail in Refs. \cite{Gross3b,SG}. We will
show how to use the subamplitudes ${T}^{ii'}_{jj'}$ in constructing
the matrix elements in a forthcoming paper \cite{prep}. 

The equations for  ${T}^{ii'}_{jj'}$ were first derived in
\cite{Gross3b} in terms of the two-body scattering
matrices ${M}^{i}_{jj'}$ by analysis of ladder and crossed ladder
exchanges. Formally they follow  from the three-body
equation in the Bethe-Salpeter framework by appropriate restriction
of the propagation of the on-shell particles. Thus, equations
for ${T}^{ii}_{jj}$ can be obtained from
Eq.~(\ref{sym4}) by replacing the two-body propagator
$G^i_{BS}$ by $G^i_j$, where $G^i_j={G}_k{\cal Q}_j$.  The
inverse propagator in the inhomogeneous term, whose role was to
cancel one of the spectator propagators when the equations are
iterated, must be replaced by unity,
$iG^{-1}_i\to 1$, since the spectator is always on shell.
This gives the following equations for ${T}^{ii}_{jj}$
\begin{eqnarray}
{T}^{ii}_{jj}=&&{1\over3}{M}^{i}_{jj}
-2\zeta\,
{M}^{i}_{jj}G^i_{j} {\cal P}_{ij} {T}^{i i}_{jj} \label{gross5}\\
=&& {1\over3}{M}^{i}_{jj}
-2\zeta\, {T}^{i i}_{jj} {\cal P}_{ij} G^i_{j} {M}^{i}_{jj} 
\label{gross6}
\end{eqnarray}
where the two body scattering operator ${M}^{i}_{jj}={\cal Q}_j
{M}^{i}{\cal Q}_{j}$ satisfies  
\begin{eqnarray}
{M}^{i}_{jj}=&&\overline{V}^{\,i}_{jj}-
\overline{V}^{\,i}_{jj} G^i_{j}{M}^{i}_{j j}\label{gross4}\\
{M}^{i}_{jj}=&&\overline{V}^{\,i}_{jj}-
{M}^{i}_{jj} G^i_{j}\overline{V}^{\,i}_{j j}\label{gross3} \, ,
\end{eqnarray}
Eq.~(\ref{gross5}) is illustrated diagrammatically in
Fig.~\ref{Gross}.   In terms of the kernels
$\overline{V}^{\,i}_{jj}$ this equation becomes
\begin{eqnarray}
{T}^{ii}_{jj}=&&{1\over3}\overline{V}^{\,i}_{jj}
-\overline{V}^{\,i}_{jj} G^i_{j}\left(1+2\zeta\, {\cal P}_{ij} \right)
{T}^{i i}_{jj}\label{gross7}\\
=&& {1\over3}\overline{V}^{\,i}_{jj}
-{T}^{i i}_{jj}\left(1+2\zeta\, {\cal P}_{ij}\right) G^i_{j}
\overline{V}^{\,i}_{jj}  \, .
\label{gross8}
\end{eqnarray}

Subvertex functions acquire an additional index to denote the
second  on-shell particle, $|\Gamma^i\rangle\rightarrow
|\Gamma^i_j\rangle={\cal Q}_i{\cal Q}_j|\Gamma^i\rangle$,   and
the symmetrized subvertex functions satisfy the following
equations
\begin{eqnarray}
|\Gamma^i_j\rangle=&&-2\zeta\,
{M}^{i}_{jj}G^i_{j} {\cal P}_{ij} |\Gamma^i_j\rangle \label{gammair}\\
=&&- \overline{V}^{\,i}_{jj}G^i_{j}
\left(1+2\zeta\,{\cal P}_{ij} \right)|\Gamma^i_j\rangle\, .
\label{gammair2}
\end{eqnarray}

Derivation of the normalization condition proceeds as for the
Bethe-Salpeter case, but for clarity we will present the full
derivation. Obtaining $\overline{V}^{\,i}_{jj}$ from
Eq.~(\ref{gross8}), and substituting this into Eq.~(\ref{gross7})
gives the following nonlinear equation for ${T}^{ii}_{jj}$
\begin{equation}
{T}^{ii}_{jj}={1\over3}\overline{V}^{\,i}_{jj}
-3T^{ii}_{jj} G^i_{j}{\cal S}_{ij}
{T}^{i i}_{jj} -3T^{ii}_{jj} {\cal S}_{ij}
G^i_{j}\,\overline{V}^{\,i}_{jj}G^i_{j}
{\cal S}_{ij} {T}^{i i}_{jj}\, , \label{nonlin}
\end{equation}
where ${\cal S}_{ij}=\left(1+2\zeta\,
{\cal P}_{ij} \right)$.
The existence of the bound state implies that
\begin{equation}
{T}^{ii}_{jj}=-{|\Gamma^i_j\rangle\,\langle \Gamma^i_j|\over M^2-P^2}
+{R}^{ii}_{jj} \, , \label {nonlin1}
\end{equation}
where, as before, ${R}^{ii}_{jj}$ is regular at the pole at $P^2=M^2$.
Requiring the coefficient of the double pole on the rhs of this
equation to be zero gives the bound state equation (\ref{gammair2}).
Equating the coefficients of the single pole terms gives an equation
similar to (\ref{2bodyeq046}):
\begin{eqnarray}
|\Gamma^i_j\rangle\,\langle{\Gamma}^{i}_j|=&&3
\lim_{P^2\rightarrow M^2}
\Biggl\{|\Gamma^i_j\rangle\;
\biggl[\frac{\langle{\Gamma}^{i}_j|\,
\left(1 +{\cal S}_{ij}G^i_{j}\,\overline{V}^{\,i}_{jj}\right)
G^i_{j}{\cal S}_{ij}
|\Gamma^i_j\rangle}{M^2-P^2}\biggr]
\;\langle{\Gamma}^{i}_j| \nonumber\\
&& -{\cal R}^{ii}_{jj}\,{\cal S}_{ij}
G^i_{j}\,\left(1 + \overline{V}^{\,i}_{jj}
G^i_{j}{\cal S}_{ij}\right)\, |\Gamma^i_j\rangle\,
\langle{\Gamma}^{i}_j|\nonumber\\
&&
 -|\Gamma^i_j\rangle\,\langle{\Gamma}^{i}_j|
\left(1 +{\cal S}_{ij} G^i_{j}\,\overline{V}^{\,i}_{jj}\right)\,
G^i_{j}{\cal S}_{ij}\,{\cal R}^{ii}_{jj}
\Biggr\}\, ,\label{2bodyeq046x}
\end{eqnarray}
where we used the fact that ${\cal S}_{ij}$ commutes with 
${\cal Q}_i G^i_{j}$.
The bound state equation ensures that the terms involving ${\cal
R}^{ii}_{jj}$ and any terms coming from the derivatives of
the subvertex functions $|\Gamma^i_j\rangle$ with respect to $P^2$ are
zero.  Since $\partial {\cal S}_{ij}/\partial P^2$ is also zero, the
condition (\ref{2bodyeq046x}) reduces to
\begin{equation}
1=-3\left\{\langle{\Gamma}^i_j|{\cal S}_{ij}
{\partial G^i_j\over \partial P^2}|\Gamma^i_j\rangle
+\langle{\Gamma}^i_j|{\cal S}_{ij} {\partial\over
\partial P^2}\left( G^i_j\overline{V}^{\,i}_{jj} G^i_j \right)
{\cal S}_{ij} |\Gamma^i_j\rangle\right\}\, .
\end{equation}
Distributing the derivative over the second term and using the bound
state equation gives our final result for the normalization condition
for the subamplitude  $| \Gamma^i_j \rangle$
\begin{equation}
1=3\left\{\langle{\Gamma}^i_j|\left(1+2\zeta {\cal
P}_{ij}\right) \left(G^i_j\right)'|\Gamma^i_j\rangle
-\langle{\Gamma}^i_j|\left(1+2\zeta {\cal
P}_{ij} \right)G^i_j
\left(\overline{V}^{\,i}_{jj}\right)' G^i_j
\left(1+2\zeta {\cal P}_{ij}\right)|\Gamma^i_j\rangle\right\}\,
.  \label{threebodygross}
\end{equation}
As in the two body case, this equation could have been
obtained directly from Eq.~(\ref{3bodyeq08b}).  

It is interesting to compare our principal result
(\ref{threebodygross}) with the nonrelativistic normalization
condition.  If $|\Psi\rangle=\sum_i|\psi^i\rangle$ is the total
three body wave function, and we choose to express the
normalization in terms of the $i=1$ component, the normalization
condition is
\begin{eqnarray}
1=&& \langle \Psi|\Psi\rangle = \langle \Psi|\left(1+\zeta
{\cal P}_{ij} + \zeta{\cal P}_{ik}\right) | \psi^i\rangle
\nonumber\\
=&& 3 \langle \Psi| \psi^i\rangle = 3 \langle \psi^i|\left(1+
\zeta{\cal P}_{ij} +\zeta{\cal P}_{ik}\right) |
\psi^i\rangle\nonumber\\ =&& 3 \langle \psi^i| \left(1+ 2\zeta{\cal
P}_{ij} \right) |\psi^i\rangle= 3 \langle \psi^1| \left(1+
2\zeta{\cal P}_{12} \right) |\psi^1\rangle \, .
\end{eqnarray}

This is very similar to the normalization condition
for the Gross vertex function, Eq.~(\ref{threebodygross}).
Adopting the convention that $i=1$ and $j=2$, using the fermion
propagator given in Eq.~(\ref{def1}), with
$k_3=P-k_1-k_2$, gives (in the rest system)
\begin{equation}
\left[G^1_2(k_3)\right]'=G'_3(k_3)=G_3(k_3)\,
{\not\! P\over 2P^2}
\, G_3(k_3) = {1\over 2M_B}\, G_3(k_3)\;\gamma^0 \; G_3(k_3) .
\end{equation}
Neglecting the terms involving the derivative of the kernel in
Eq.~(\ref{threebodygross}), decomposing the propagator into
positive and negative energy parts,
\begin{equation}
G_3(k_3)=\left({1\over2 E_{k_3}}\right)\sum_\lambda\left[
{u({\bf k}_3,\lambda)\bar{u}({\bf k}_3,\lambda)\over E_{k_1}
+E_{k_2}+E_{k_3}-M_B}-
{v(-{\bf k}_3,\lambda)\bar{v}(-{\bf k}_3,\lambda)\over
E_{k_3}+M_B-E_{k_1}-E_{k_2}}\right]\, ,
\end{equation}
keeping only the positive energy part and  using
$\bar{u}({\bf k}_3,\lambda')\gamma^0{u}({\bf k}_3,\lambda) =
2E_{k_3}\delta_{\lambda'\lambda}$, allows us to reduce the
normalization condition (\ref{threebodygross}) for the bound state
vertex function for three identical fermions to
\begin{equation}
1\simeq 3\sum_\lambda \langle\psi^1_\lambda
|\left(1+2\zeta {\cal P}_{12}\right)
|\psi^1_\lambda\rangle \, ,
\end{equation}
where the wave function is related to the vertex function by
\begin{equation}
|\psi^1_\lambda\rangle \simeq {1\over \sqrt{2M_B}}\;
{u({\bf k}_3,\lambda) |\Gamma^1_2\rangle\over E_{k_1}
+E_{k_2}+E_{k_3}-M_B}\, .
\end{equation}
This correspondence will be developed in greater detail elsewhere.
Here we only wish to emphasize that the covariant normalization
condition (\ref{threebodygross}) reduces to the expected
nonrelativistic limit.

\section{CONCLUSION}

The principal results of this paper are the derivation of the
normalization conditions (\ref{3bodyeq07}) and
(\ref{3bodyeq08b}) for the three-body BS vertex functions, and
(\ref{threebodygross}) for the three-body Gross vertex
functions.  These results are new.

In the course of deriving these results we also showed that the
normalization condition for the two-body Gross equation is
identical to the requirement that the charge of a bound state
be equal to sum of the charges of its constituents.  Since we
derived the normalization condition without any reference to the
charge, this is equivalent to a proof that charge is
automatically conserved by the spectator theory.  This
result has been used in previous work \cite{deutletter},
but the proof is new.   A similar proof should be
possible for the three body system, and will be presented
in a subsequent  paper \cite{prep}.  
 
\acknowledgments

We are happy to acknowledge the support of the DOE through
Jefferson Laboratory and grant No. DE-FG05-88ER40435.

\appendix

\section*{Feynman Rules}

In this paper, we use the Feynman rules as given in
\cite{GROSS1}.

The variation from the usual field theoretical practice is with regard to
the normalization of the symmetrized two- and three-body scattering matrices
${M}$ and ${T}$. The practice in most field theory
texts is to  use an unnormalized symmetrization operator
and to include all information  about the symmetry of the
scattering process in an overall phase space  factor.
Here we choose to use normalized symmetrization operators
${\cal A}_2$ for the two-body case and ${\cal A}_3$ for
the three-body case, as defined by (\ref{symmetry3}). This
results in integral equations for the scattering
amplitudes which are consistently of the form of the
nonrelativistic  Lippmann-Schwinger equations and
preserves the nonrelativistic normalization  of the wave
functions. Any matrix element that involves an occurrence
of the  symmetrized scattering matrices as used here can
be used in the usual  expression for cross sections by
making the substitutions
${M}\rightarrow 2{ M}$ and ${T}\rightarrow
6{T}$.

%\end{document}

\pagebreak

\centerline{ FIGURE CAPTIONS}

\vspace*{0.8truecm}

%figure1  
%2_body_scatt_mat.fig.eps
Figure~1: Diagrammatic representation of the two-body BS equation
for the scattering matrix.

\vspace*{0.8truecm} 

%figure2
% symmetrized.fig.eps
Figure~2: Diagrammatic representation of the symmetrized kernel.

\vspace*{0.8truecm} 
 
%figure3 
% 2_body_scatt_mat_gr.fig.eps
Figure~3: Diagrammatic representation of the two-body Gross equation
for the scattering matrix.  The $\times$ on the line for particle 1 indicates
that it is on shell.
 
\vspace*{0.8truecm} 

%figure4 
% symmetrized_gr.fig.eps
Figure~4: Diagrammatic representation of the symmetrized kernel for Gross
equation.  As in Fig.~3, the $\times$ indicates that 
the particle is on shell.

\vspace*{0.8truecm} 
 
%figure5 
% 2_body_gross_norm.fig.eps
Figure~5: Diagrammatic representation of the normalization condition for the
two-body Gross vertex function.  The dashed line represents the derivative
$\partial/\partial P^2$, and the $\times$ means that the particle is on shell.

\vspace*{0.8truecm} 
  
%figure6 
% 2_body_gross_formf.fig.eps
Figure~6: Diagrams which contribute to the electromagnetic charge of the two
body bound state.

\vspace*{0.8truecm} 

%figure7 
% 2_body_poles.fig.eps
Figure~7: Location in the $p_{10}$ complex plane of the 6 poles from the three
nucleon propagators in Eq.~(2.34).  The last two terms in Eq.~(2.33) emerge if
the integral of $p_{10}$ over the contour $C$ enclosing the poles 3a and 1a is
evaluated using the residue theorem.

\vspace*{0.8truecm} 
  
%figure8  
% faddeev.fig.eps
Figure~8: Diagrammatic representation of the Faddeev equations for the amplitudes
${\cal T}^{1i}$.  Note that the spectator is identified by the solid dot.

\vspace*{0.8truecm}

%figure9  
%gross.fig.eps
Figure~9: Diagrammatic representation of the Gross equation for the amplitude
$T^{11}_{22}$.  On-shell particles are labeled with an $\times$.  The final
state particles emerge from the left side of each diagram, and 
putting the final state spectator on-shell consistently  automatically 
forces a second particle in the final state to be on-shell.  In this way 
two particles are always on-shell.
 
\end{document}